\documentclass[11pt,a4paper]{article}

\usepackage{indentfirst}
\usepackage{amsfonts}
\usepackage{amssymb}
\usepackage[reqno,intlimits]{amsmath}
\usepackage[polish,english]{babel}

\newtheorem{theorem}{Theorem}
\newtheorem{lemma}{Lemma}
\newtheorem{conjecture}{Conjecture}[theorem]

\newcommand{\ud}{\mathrm{d}}
\newcommand{\rmd}{\mathrm{d}}
\newcommand{\R}{\mathbb{R}}
\newcommand{\N}{\mathbb{N}}
\newcommand{\C}{\mathbb{C}}
\title{Global integrability  of cosmological scalar fields}
\author{Andrzej J.~Maciejewski,\\
Institute of Astronomy,
University of Zielona G\'ora\\
Podg\'orna 50, 65--246 Zielona G\'ora, Poland,
(e-mail: maciejka@astro.ia.uz.zgora.pl)\\[2ex]
Maria Przybylska,\\
Toru\'n Centre for Astronomy,
Nicholaus Copernicus University \\
Gagarina 11, 87--100 Toru\'n, Poland, (e-mail:
mprzyb@astri.uni.torun.pl)\\[2ex]
Tomasz Stachowiak,\\
Astronomical Observatory, Jagiellonian University\\
Orla 171, 30-244 Krak\'ow, Poland, (e-mail:
toms@oa.uj.edu.pl)\\[2ex]
Marek Szyd{\l}owski,\\
Marc Kac Complex Systems Research Center, Jagiellonian University\\
Reymonta 4, 30-059 Krak{\'o}w, Poland\\
and\\
Astronomical Observatory, Jagiellonian University\\
Orla 171, 30-244 Krak\'ow, Poland, (e-mail:
uoszydlo@cyf-kr.edu.pl)}

\begin{document}
\maketitle

\begin{abstract}
We investigate the Liouvillian integrability of Hamiltonian systems
describing a universe filled with a scalar field (possibly complex). The tool
used is the differential Galois group approach, as introduced by
Morales-Ruiz and Ramis. The main result is that the generic systems with
minimal coupling are non-integrable, although there still exist some values of
parameters for which integrability remains undecided; the conformally coupled
systems are only integrable in four known cases. We also draw a connection with the
chaos present in such cosmological models, and the issues of the integrability
restricted to the real domain.
\end{abstract}

\section{Introduction}

Homogeneous and isotropic cosmological models, although very simple,
explain the recent observational data very well
\cite{Riess:1998cb,Perlmutter:1998np}. Their foundation
is the Friedmann-Robertson-Walker (FRW) universe, described by the metric
\begin{equation}
    \ud s^2 = a(\eta)^2 \left[ - \ud\eta^2 + \frac{\ud r^2}{1-Kr^2} +
    r^2\ud^2\Omega_2 \right],\label{metric}
\end{equation}
where $a$ is the scale factor, $\ud^2\Omega_2$ is the line element on a
two-sphere, and we chose to use the conformal time $\eta$. As can be seen from
the above metric, the scale factor represents the relative change in the
distance of two points whose spatial coordinates are fixed. It depends only on
time, so that the whole universe is deformed in a homogeneous fashion. From the
anthropocentric point of view it could be seen as a three dimensional space
evolving in the time - in the simplest case when the curvature index $K$ is zero,
it would be a Euclidean space stretched according to~$a$.

If we were to fill such a universe with a matter, its properties could only
depend on time, and be the same in all points of the spatial subspace at a
given value of $\eta$ -- otherwise the model would no longer be homogeneous.

For example a perfect fluid would be completely described by two quantities
-- its density and pressure as functions of time. A scalar field would be
described by its field variables. Also a cosmological constant with its trivial
dependence on time could always be included in such models.

Depending on the matter components one obtains various evolutions of the scale
factor $a$, as given by the general action
\begin{equation}
    \mathcal I = \frac{c^4}{16\pi G}\int\left[ \mathcal R - 2\Lambda -
    \frac12\left(\nabla_{\alpha}\bar{\psi}\nabla^{\alpha}\psi+
    V(\psi)+\xi\mathcal R|\psi|^2\right) -\varrho
    \right]\sqrt{-g}\;\ud^4\mathrm{x},
\label{gen_act}
\end{equation}
where $\mathcal R$ is the Ricci scalar, $\Lambda$ the cosmological constant,
$V$ the field's potential, $\xi$ the coupling constant, 
and $\varrho$ is the density of the perfect fluid. The potential usually
includes at least a quadratic term $m^2|\psi|^2$, where $m$ is the so-called mass
of the field. When $\xi=0$ we say that the
field is minimally coupled -- it does not uncouple since the determinant of the
metric $g$ multiplies the whole Lagrangian density. Case with $\xi=\frac16$ is the so-called conformal coupling.

For the considered geometry, the above action can be simplified so that it
allows the Hamiltonian approach with the phase variables depending only on
conformal time $\eta$. Due to the required covariance of general relativity,
the system is subject to constraints, which in our case mean that the obtained
Hamiltonian's value is zero. However, we note that including an additional matter
component $\varrho$ is equivalent to considering other energy levels. Namely
for $\varrho\propto a^{-4}$ (which is the case for radiation) a constant is
added to the Hamiltonian, thus imitating its non-zero value. This is the
justification for studying the systems integrability on a generic energy
hyper-surface.

From the observational point of view, the cosmological constant provides an
explanation for the current accelerating
expansion of the universe ($\Lambda$ Cold Dark Matter model
\cite{Copeland:2006wr}), but a better solution still is sought for. A real scalar
field dubbed ``quintessence'' with the so-called slow rolling
potential, which models the dark energy component has been extensively used for
that purpose \cite{Caldwell:1997ii,Zlatev:1998tr}. Realisations of the field
itself include also Bose-Einstein
condensate of axions \cite{Dymnikova:2001ga} or a phantom violating the energy
principle \cite{Caldwell:1999ew}.

Finally, scalar field could also be the
mechanism behind the inflation \cite{Linde:1981mu,Lidsey:1995np}, which is
currently the most established and used scenario for the early Universe
\cite{Boyanovsky:2005pw}.
Recently, Komatsu et al. \cite{Komatsu:2008hk} have shown that latest observational
data (WMAP, SNIa, Barion Oscillation Peak and others) show that the model of
chaotic inflation (which strictly speaking should be called non-integrable or
complex in the sense that we demonstrate in the present work) with the
quadratic potential remains a good fit (within the 95\% confidence domain).

From the physical point of view, a universe filled with only one component
seems simple enough but it is not the case here. Chaotic
scattering has been found in minimally coupled fields \cite{Cornish:1997ah},
as has chaotic dynamics \cite{Monerat:1998tw}.

The first of our results is that minimally coupled fields are not integrable in
the generic case. There are however special families of the system's parameters
which leave the question open. We give the appropriate conditions in the
concluding section.

There are several physical reasons to study more than just minimally coupled
fields. Early works on chaotic inflation found the coupling constant
$\xi$ small or negative \cite{Futamase:1989} but some argue \cite{Faraoni:2000gx}
that the paradigm of inflation should be generalised to the case with non-zero
coupling constant which should not be fine-tuned close to zero, and WMAP
observations seem to indicate non-negligible $\xi$.

The coupling could be generated by quantum corrections
\cite{Birrel,Ford:1981xj}, or from the renormalisation of the Klein-Gordon
equation as described in
\cite{Callan:1970ze}. The coupling constant should be fixed by particle physics
of the matter composing the scalar field, for example the way $\xi=1/6$ was
found in the large $N$ approximation to the Nambu-Jona-Lasimo model in
\cite{Hill:1991jc}. Non-minimally coupled fields are also interesting in the
context of description of the dark energy for which the ratio between the pressure and
the energy density is less than $-1$. Such matter is called a phantom matter, and
cannot be achieved by standard scalar fields \cite{Faraoni:2001tq}.

Conformally coupled fields were subject to more rigorous integrability
analysis, as opposed to minimally coupled ones, thanks to the natural form of
their Hamiltonian. As will be shown in the next section, the kinetic part is of
natural form, albeit indefinite, and the potential is polynomial (in the case
of real fields). 

Chaos has been studied in such fields by means of
Lyapunov exponents, perturbative approach, breaking up of the KAM tori
\cite{Calzetta:1992bv,Bombelli:1997ut}. Also
the Painlev\'e property \cite{Helmi:1997mj} was employed as an indicator of the
system's integrability.

Ziglin proved that the system given by (\ref{ham_orig1}) is not meromorphically
integrable when $\Lambda=\lambda=0$ and $k=1$ \cite{Ziglin:00::}. His methods
were also used by Yoshida to homogeneous potentials which is the case for the
system when $k=0$ \cite{Yoshida:86::,Yoshida:87::,Yoshida:88::,Yoshida:89::}.
Later, Yoshida's results were sharpened by Morales-Ruiz and Ramis
\cite{Morales:99::}, and used by the present authors in
\cite{Maciejewski:00::b} to obtain countable families of possibly integrable cases with some restrictions on
$\lambda$ and $\Lambda$. Also recently,
more conditions for integrability have been given in \cite{Boucher}, although
only for a non-zero spatial curvature $k$ and a generic value of energy, that is,
when the particular solution is a non-degenerate elliptic function defined 
on a non-zero energy level.

Our work shows, that the conjecture of that paper is in fact correct -- as shown
in Section 4 -- the conformal system is only integrable in two cases (with the above
assumptions). We go further than that
and show that for a generic energy value, a spatially flat ($k=0$) the universe is
only integrable in four cases. Also, the particular case of zero energy is
analysed and new, simple conditions on the model parameters are found. Finally,
we check that when $E=k=0$, the problem remains open, as the necessary
conditions for integrability are fulfilled.

When it comes to numerical studies of the problem, there are various results,
most notably a chaotic
behaviour \cite{Joras:2003dn}, but also a fractal structure and chaotic scattering
\cite{Szydlowski:2006qn}. However, it remains unclear whether the widely
exercised complex rotation of the variables changes the system's integrability.
Even for very simple systems it was shown \cite{Gorni} that there might exist
smooth integrals, which are not real-analytic. This question is especially
vital since our Universe clearly has real initial conditions and dynamics.

The results obtained for the conformal coupling are much stronger than in the
minimal one. We manage to show, that the four cases with known first integrals
are the only integrable ones for the generic energy value.

The Hamiltonian of both these systems has indefinite kinetic energy part, and to
cast it into a positive-definite form a transition into imaginary variables is
used. It has been done for a conformally coupled field \cite{Joras:2003dn},
but some authors, see e.g.  \cite{Motter:2002jj},  argue that  there are physical limitations
which forbid extending the solutions through singularities such as $a=0$, and
an imaginary scale factor seems even less realistic.

We would like to stress that the complexification of the variables in our
approach is not to be connected with the physical evolution of the system into the
imaginary values. The behaviour of differential equations in the complex domain
is a tool that allows for obtaining general results regarding its
integrability (also real), as can be seen on the example of the Painlev\'e analysis
\cite{Helmi:1997mj}.

Despite the fact, that systems of the considered type were often called
non-integrable, there was seldom a rigorous proof of that proposition. However,
the Liouvillian integrability can be studied successfully, as we try to show in
this article.

The authors are aware of one notable attempt at studying the
problem in \cite{Coelho:2008}; sadly, that paper contains a serious mistake in the
application of Yoshida's theorem. The method used requires rescaling by the
energy of the system, which the authors of \cite{Coelho:2008} take to be zero.
Thus, their
Theorem~3 (which contradicts one of the results presented here) is in fact false.
The zero energy level usually requires a separate treatment, which we
also present here.

The plan of the paper is the following. In the next section we describe the mentioned two cosmological models: of minimally coupled field and conformally coupled field starting from a general description up to the Hamiltonian formulation of these models. Section~\ref{sec:theory} is devoted to introduction to 
 the Morales-Ramis theory. This is our tool to prove the announced rigorous integrability results for these models. In the next two Sections \ref{sec:minimally} and \ref{sec:conformally} all details of integrability analysis are presented. For convenience of readers all integrability results are recapitulated in Section~\ref{sec:conclusions}.

\section{Physical system's setup}
\label{sec:setup}
\subsection{Minimally coupled field}

The general action (\ref{gen_act}) now includes the following parts
\begin{equation}
    \mathcal{I} = \frac{c^4}{16\pi G}\int \left[\mathcal{R} - 2\Lambda -
    \frac12\left(\nabla_{\alpha}\bar{\psi}\nabla^{\alpha}\psi +
    \frac{m^2}{\hbar^2}|\psi|^2\right)\right]\sqrt{-g}\,\ud^4\boldsymbol{\rm x},
\end{equation}
Using the coordinates of (\ref{metric}), the Lagrangian becomes
\begin{equation}
    \mathcal L = 6(\ddot a a + Ka^2) + \frac12 |\dot\psi|^2a^2 -
    \frac{m^2}{2\hbar^2}a^4|\psi|^2 - 2\Lambda a^4,
\end{equation}
with the dot denoting the derivative with respect to time.
We also dropped a coefficient which includes some physical
constants and the part of the action related to the spatial integration.

Next we subtract a full derivative $6(\dot{a}a)\dot{}$, and use the polar parametrisation
for the scalar field $\psi = \sqrt{12}\,\phi \exp(i\theta)$ to get
\begin{equation}
    \mathcal L = 6(Ka^2 - \dot a^2) + 6a^2(\dot\phi^2+\phi^2\dot\theta^2) -
    6\frac{m^2}{\hbar^2}a^4\phi^2- 2\Lambda a^4,
\end{equation}
and obtain the Hamiltonian
\begin{equation}
    H = \frac{1}{24}\left(\frac{1}{a^2}p_{\phi}^2 - p_a^2 +
    \frac{1}{a^2\phi^2}p_{\theta}^2\right) - 6Ka^2 + 2\Lambda a^4 +
    6\frac{m^2}{\hbar^2}a^4\phi^2,
\end{equation}
with
\begin{equation}
    p_a = -12\dot{a},\quad p_{\phi} = 12a^2\dot\phi,\quad p_{\theta} =
    12a^2\phi^2\dot\theta.
\end{equation}
Note that the the elliptic constraints of general relativity require that the
above Hamiltonian is identically zero -- this is the so-called Friedmann
equation, although it is not a dynamical evolution equation but rather a
conservation law \cite{Misner}.

Since $\theta$ is a cyclical variable, the corresponding momentum is
conserved so we substitute $p_{\theta}^2=2\omega^2$. To make all the
quantities dimensionless, we make the following rescaling
\begin{equation}
    m^2\rightarrow m^2\hbar^2|K|,\quad
    \Lambda\rightarrow 3\Lambda|K|,\quad \omega^2\rightarrow72\omega^2|K|,
    \quad p_a^2\rightarrow 72p_a^2|K|,\quad p_{\phi}^2\rightarrow
    72p_{\phi}^2|K|.
\end{equation}
There is no need of changing the variables $a$ and $\phi$ along with their
momenta, as this is really changing the time variable $\eta$, and thus the
derivatives to which the momenta are proportional. This results also in
dividing the whole Hamiltonian by $6\sqrt2|K|$ to yield
\begin{equation}
    \sqrt2 H = \frac12 \left(-p_a^2 + \frac{1}{a^2}p_{\phi}^2 \right) -
    \frac{K}{|K|} a^2 + \Lambda a^4
    + m^2\phi^2 a^4 +\frac{\omega^2}{a^2\phi^2}.
\end{equation}
If the spatial curvature is zero, any of the other dimensional constants can be
used for this purpose, so
without the loss of generality we take the right-hand side to be the new
Hamiltonian
\begin{equation}
    H = \frac12 \left(-p_a^2 + \frac{1}{a^2}p_{\phi}^2 \right) - k a^2 + \Lambda a^4
    + m^2\phi^2 a^4 +\frac{\omega^2}{a^2\phi^2}, \label{MainHam}
\end{equation}
and in all physical cases, $k\in\{-1,0,1\}$, $\omega^2\geqslant 0$,
$m^2\in\mathbb R$, $\Lambda\in\mathbb R$, $H=0$. We extend the analysis somewhat
assuming that the Hamiltonian might be equal to some non-zero constant
$E\in\mathbb R$. We will later see, that our analysis
includes also the possibility of these coefficients being complex.

Note that for a massless ($m=0$) field, the system is already solvable, as
shown in Appendix~A.

From this point on, we take $\omega=0$, which means the phase is constant.
Since the model has U(1) symmetry, we can always make such a field real with a
rotation in the complex $\psi$ plane. In other words, we will be investigating
a real scalar field only. The reason why we restrict the problem is the
following: the method employed requires an explicit (non-constant) particular
solution, and the only one known requires $\phi=0$; which is a singularity of
the full Hamiltonian.

Under the above assumption the Hamilton's equations of system (\ref{MainHam}) are
\begin{equation}
\begin{aligned}
    \dot{a} &= -p_a, &&&
    \dot{p}_a &= 2k a - 4a^3(\Lambda +m^2\phi^2) + \frac{1}{a^3}p_{\phi}^2,\\
    \dot{\phi} &= \frac{1}{a^2}p_{\phi}, &&&
    \dot{p}_{\phi} &= -2m^2 a^4\phi.
\end{aligned} \label{MainEq}
\end{equation}

We note that there is an obvious particular solution, which describes an empty
universe: $\phi=p_{\phi}=0$, $a=q$, $p_a=-\dot{q}$. Thanks to the energy integral
$E=\frac12\dot{q}^2+kq^2-\Lambda q^4$, it can be
identified with an appropriate elliptic function.

\subsection{Conformally coupled scalar fields}

The procedure of obtaining the Hamiltonian is the same as in the case of
minimally coupled fields, only this time the action is
\begin{equation}
    \mathcal{I} = \frac{c^4}{16\pi G}\int \left[\mathcal{R} - 2\Lambda -
    \frac12\left(\nabla_{\alpha}\bar{\psi}\nabla^{\alpha}\psi +
    \frac{m^2}{\hbar^2}|\psi|^2 + \frac16\mathcal{R}|\psi|^2\right)
    - \frac{\lambda}{4!}|\psi|^4\right]\sqrt{-g}\,\ud^4\boldsymbol{\rm x},
\end{equation}
where an additional coupling to gravity through the Ricci scalar $\mathcal R$,
and a quartic potential term with constant $\lambda$
are present, as opposed to the minimal scenario. We keep the
same notation as before and express the involved quantities in the same
coordinates and get
\begin{equation}
    \mathcal L = 6(\ddot{a}a + Ka^2) - \frac12\ddot{a}a|\psi|^2 + \frac12
    |\dot{\psi}|^2a^2 -
    \frac{m^2}{2\hbar^2}a^4|\psi|^2 -\frac{\lambda}{4!}a^4|\psi|^4 - 2\Lambda
    a^4 - \frac12K a^2|\psi|^2,
\end{equation}
from which we remove a full derivative, and introduce new field variables $\psi
= \sqrt{12}\,\phi\exp(i\theta)/a$ to obtain
\begin{equation}
    \mathcal L = 6\left[ \dot\phi^2 + \phi^2\dot\theta^2 - \dot{a}^2 + K(a^2-\phi^2)
    -\frac{m^2}{\hbar^2}a^2\phi^2 - \frac{\Lambda}{3}a^4 - \lambda\phi^4\right].
\end{equation}
The associated Hamiltonian is
\begin{equation}
    H = \frac{1}{24}\left(p_{\phi}^2 +\frac{1}{\phi^2}p_{\theta}^2 - p_a^2\right)
    + 6\left[ K(\phi^2-a^2) + \frac{m^2}{\hbar^2}a^2\phi^2 + \lambda\phi^4 +
    \frac{\Lambda}{3} a^4 \right],
\end{equation}
with
\begin{equation}
    p_a = -12\dot{a},\quad p_{\phi} = 12\dot\phi,\quad p_{\theta} =
    12\phi^2\dot\theta.
\end{equation}
We can see that $\theta$ is a cyclical variable because we took the potential
to depend on the modulus of $\psi$ only, so we write a constant instead of the
respective momentum $p_{\theta}=\omega$.

Finally, we express everything in dimensionless quantities, rescaling the
constants, but also the time and momenta (as they are in fact time
derivatives), which results in rescaling the whole Hamiltonian. We do this as
follows
\begin{equation}
    m^2\rightarrow m^2\hbar^2|K|,\quad \Lambda\rightarrow \frac32\Lambda|K|,
    \quad \lambda\rightarrow\frac12\lambda|K|,
    \quad p_x^2\rightarrow 144 p_x^2 |K|,
    \quad H\rightarrow\frac{1}{12\sqrt{|K|}}H,
\end{equation}
when $K\ne0$, and using another of the dimensional constants otherwise.
Thus, eliminating a multiplicative constant, the Hamiltonian reads
\begin{equation}
    H = \frac12\left(p_{\phi}^2 - p_a^2\right) + \frac12\left[k(\phi^2 - a^2) +
    \frac{\omega^2}{\phi^2} + m^2 a^2\phi^2\right] +\frac14\left(\Lambda a^4 +
    \lambda\phi^4\right), \label{ham_orig}
\end{equation}
with $k\in\{-1,0,1\}$ ($K=k|K|$); $\omega$, $\lambda$, $\Lambda$, $m^2
\in\mathbb R$, and $H=0$
in any physically possible setup. Exactly as in the previous case, the zero
value of the energy is a consequence of the constraints introduced by general
relativity.

We note that for $m=0$ the system decouples, and is trivially integrable as
shown in Appendix~B. That is why we will assume $m\ne0$ henceforth. We will
also take $\omega=0$, that is, consider a scalar field equivalent to a real
field after a unitary rotation in the complex $\psi$ plane.

We change the field variables into the standard $q$ and $p$ ones for further
computation, taking
\begin{equation}
\begin{aligned}
    a &= q_1,&\quad p_a&=p_1,\\
    \phi &= q_2,&\quad p_{\phi}&=p_2.
\end{aligned}
\end{equation}
The Hamiltonian is then
\begin{equation}
\begin{aligned}
    H &= \frac12\left(-p_1^2 + p_2^2\right) + V,\\
    V &= \frac12\left[k(-q_1^2 + q_2^2) 
    + m^2 q_1^2q_2^2\right] +\frac14\left(\Lambda q_1^4 +\lambda q_2^4\right).
\end{aligned}\label{ham_orig1}
\end{equation}

\section{Differential Galois obstructions  to integrability}
\label{sec:theory}

Let $(M,\omega)$ be a $2n$-dimensional complex analytic symplectic manifold. For a meromorphic function $H:M\rightarrow \mathbb{C}$, we denote by  $V_H$  the Hamiltonian vector field generated by $H$ and let us consider  Hamiltonian equations
\begin{equation}
\dfrac{\mathrm{d} x}{\mathrm{d}t}=v_H(x),\qquad t\in\mathbb{C},\quad x\in M.
\label{eq:dynsys}
\end{equation}
  We assume that a non-constant
particular solution~$\varphi(t)$ of system \eqref{eq:dynsys} is known.
Its maximal analytic continuation
defines a Riemann surface~$\Gamma$ with the local coordinate~$t$.  

Linearisation of \eqref{eq:dynsys}
around $\varphi(t)$ 
yields  variational equations of the following  form
\begin{equation}
\label{eq:vds}
 \dot{\xi} = A(t)\xi, \qquad  A(t)=\frac{\partial
v_H}{\partial x}(\varphi(t)),  \qquad \xi \in T_\Gamma M.
\end{equation}

Thanks to Hamiltonian character of the system  the dimension of variational equations can be reduced by two.
First we use the fact that a Hamiltonian system has at least one first
integral namely Hamiltonian $H$, thus we can restrict system
\eqref{eq:dynsys} to the manifold $M_\varepsilon=\{x\in
M\,|\,H(x)=\varepsilon\}$, where $\varepsilon=H(\varphi(t))$.  Then we
consider the induced system on the normal bundle $N:=T_\Gamma
M_\varepsilon /T\Gamma$ of $\Gamma$
\begin{equation}
\label{eq:nve}
 \dot \eta =\widetilde{A}(t)\eta,\qquad  \widetilde{A}(t)\eta=\pi_\star(T(v)(\pi^{-1}\xi)), \qquad \eta\in N.
\end{equation}
Here $\pi: T_\Gamma M_\varepsilon\rightarrow N$ is the projection. The
system of $2n-2$ equations obtained in this way is called the normal
variational equations.

 We can consider the entries of matrices $A$  and $\widetilde{A}$ as elements
of  field ${\mathcal K}:={\mathcal M}(\Gamma)$  of meromorphic functions  on $\Gamma$.  This
field with differentiation with respect to $t$ as a derivation is  a
differential field. Only constant functions from ${\mathcal K}$ have a vanishing derivative,
so the subfield of constants of ${\mathcal K}$ is $\mathbb{C}$.

It is obvious that  solutions of~\eqref{eq:vds} are not necessarily elements
of  ${\mathcal K}^n$. The fundamental theorem of the differential Galois theory guarantees
that there exists a differential field ${\mathcal F}\supset {\mathcal K}$ such that it contains $n$
linearly independent (over $\mathbb{C}$) solutions of~\eqref{eq:vds}. The smallest
differential extension  ${\mathcal F}\supset {\mathcal K}$  with this property  is called the
Picard-Vessiot extension. A group ${\mathcal G}$ of differential automorphisms of
${\mathcal F}$ which does not change ${\mathcal K}$ is called the differential Galois group of
equation~\eqref{eq:vds}.
It can be shown that ${\mathcal G}$ is a linear algebraic group.  Thus, it is a union
of disjoint connected components. One of them containing the identity is called
the identity component of ${\mathcal G}$. 

Differential Galois theory was created as a tool to answer the question: whether a given system of linear equations possesses a solution that can be written in a closed form, i.e. is it solvable? The main theorem of this theory states that the necessary condition of solvability in the class of Liouvillian functions (i.e. by generalised quadratures)  is solvability of its differential Galois group. We can try to connect the integrability of the original nonlinear system with solvability of its variational equations. However, there is a more direct connection. Namely in eighties of XX century  Ziglin observed that if system ~\eqref{eq:dynsys} has
$k\geq 2$ functionally independent meromorphic first integrals, then 
 variational equations \eqref{eq:vds} and also normal variational equations \eqref{eq:nve} possess $k$  rational first integrals and moreover the monodromy group (that is a subgroup of differential Galois group) has the same number of invariants  \cite{Ziglin:82::b,Ziglin:83::b}. Fourteen years later the relation between first integrals and invariants of the differential Galois group was analysed by Baider, Churchill, Rod and Singer in \cite{Churchill:96::b}. However the final formulation of relations between integrability of Hamiltonian systems and properties of the differential Galois group of variational equations due to Morales and Ramis \cite{Morales:01::b1,Morales:99::} where in their analysis not only the presence of first integrals is taken into account but also the consequences of the involution of first integrals. Their main theorem that will be the crucial tool of our analysis is the following.
\begin{theorem}[Morales-Ruiz and Ramis \cite{Morales:99::}]
  Assume that a Hamiltonian system is meromorphically integrable in
  the Liouville sense in a neighbourhood of a phase curve
  $\Gamma$ and  irregular singularities of the variational equations along $\Gamma$ do not correspond to phase points at the infinity. Then the identity component of the differential Galois group
  of the (normal) variational equations associated with $\Gamma$ is
  Abelian.
\label{thm:MR}
\end{theorem}
Let us explain assumptions concerning variational equations in the above theorem. Usually the Riemann surface corresponding to the phase curve $\Gamma$ is not compact so we compactify it adding some points. Typically these points correspond to equilibria or infinite points. In the later case we have to add these points to the phase space, i.e, we have to extend our original system into a `bigger' phase space.  For the extended system  the requirement that the considered first integrals are meromorphic in a neighbourhood of the phase curve put strong restrictions: they have to be meromorphic at the infinity.  Thus if we remove the assumption about irregular singular points we have to restrict the class of first integrals.  Below we give a version of the Morales-Ramis theorem without assumptions concerning the regularity of variational equations, which is adapted to Hamiltonian systems considered in this paper.

\begin{theorem}[Morales-Ruiz and Ramis \cite{Morales:99::}]
  Assume that a Hamiltonian system defined in a linear symplectic space is generated by a rational Hamiltonian function and is rationally integrable in
  the Liouville sense. Then the identity component of the differential Galois group
  of the (normal) variational equations associated with $\Gamma$ is
  Abelian.
\end{theorem}

In applications the most difficult part is to check the abelianity of variational equations. Fortunately, thanks to the separation of variational equations into two parts we can restrict to its normal part and in this way to reduce the dimension of the system. Furthermore, because the abelian differential Galois group implies in particular that this group is solvable, thus we can use directly all solvability results concerning some known equations such as e.g. hypergeometric equation, Whittaker equation, Lam\'e equations. In addition, for a linear second order equation with rational coefficients there exists the closed algorithm, so-called Kovacic algorithm \cite{Kovacic:86::},  that decides whether equation is solvable in a class of Liouvillian function, yields  explicit forms of solutions as well as determines the differential Galois group. This is achieved by providing necessary
conditions for solvability of the appropriate Galois group. The equations in
question have as their Galois group an algebraic subgroup of
$\mathrm{SL}_2(\mathbb C)$, and since there are only three possibilities of
those having a solvable identity component, the procedure is arranged in three
cases. They consist of analysing the equation's singular points and
finding an appropriate polynomial and an algebraic function of possible
degrees 2, 4, 6, or 12; used to construct solutions.

This means that  Theorem~\ref {thm:MR}  yields really an effective tool for proving the
non\--in\-teg\-ra\-bi\-li\-ty and distinguishing the cases suspected about integrability in the case when Hamiltonian depends on some physical parameters, for examples of applications see references in \cite{Morales:06::}.

It can happen that a considered system satisfies all conditions of the above
theorem, but nevertheless it is not integrable. It is nothing strange as this
theorem gives only necessary conditions for the integrability, This shows a need of stronger
necessary conditions for the integrability. They were developed by C.~Sim\'o,
J.J~Morales and J.-P.~Ramis \cite{Morales:99::,Morales:00::,Morales:06::} and are based on higher order variational equations (HVE's).

\begin{theorem}
\label{thm:basicGG}
Assume that a Hamiltonian system is meromorphically integrable in the
Liouville sense in a neighbourhood of the analytic phase curve
$\Gamma$, and the infinity is a regular singular point of the variational
equations along $\Gamma$.  Then the identity component of the differential
Galois group of $k$-th variational equations along $\Gamma$ is Abelian for all
$k\geq 1$.
\end{theorem}
For variational equations (VE's) of degree greater than one there is no more the splitting of variational equations into two parts: normal (NVE's) and tangential (TVE's) and there is no reduction of system's dimension and the analysis of the differential Galois group of the whole system of variational equations is more involved.
Fortunately, in the case when variational equations are the product of Lam\'e equations with Lam\'e-Hermite solutions Morales-Ruiz proved  in \cite{Morales:99::,Morales:00::,Morales:06::} that the absence of logarithmic terms in solutions of higher order variational equations is a necessary condition of abelianity of the identity component of their differential Galois groups.

 The interested reader  more detailed and 
complete presentation of Morales-Ramis theory can find 
in \cite{Churchill:96::b,Morales:99::,Morales:01::b1,Ziglin:82::b,Ziglin:83::b} and of differential 
Galois theory in \cite{Beukers:92::,Singer:90::,Kaplansky:76::,Morales:99::,Put:03::}.

\section{Analysis of the minimally coupled field}
\label{sec:minimally}
\subsection{$\Lambda=0$ case}
\label{sec:minimallyL0}
The system now has the following form
\begin{equation}
\begin{aligned}
    \dot{a} &= -p_a, &&&
    \dot{p}_a &= 2k a - 4m^2a^3\phi^2 + \frac{1}{a^3}p_{\phi}^2,\\
    \dot{\phi} &= \frac{1}{a^2}p_{\phi}, &&&
    \dot{p}_{\phi} &= -2m^2 a^4\phi.
\end{aligned}
\end{equation}
Using the aforementioned particular solution, for which the
constant energy condition becomes $E=\frac12\dot{q}^2+kq^2$, we have as the
variational equations
\begin{equation}
\left( \begin{array}{c}
\dot{a}^{(1)}\\
\dot{p}_a^{(1)}\\
\dot{\phi}^{(1)}\\
\dot{p}_{\phi}^{(1)}
\end{array} \right) = \left( \begin{array}{cccc}
0 & -1 & 0 & 0 \\
2k & 0 & 0 & 0 \\
0 & 0 & 0 & q^{-2} \\
0 & 0 & -2m^2q^4 & 0
\end{array} \right)
\left( \begin{array}{c}
a^{(1)}\\
p_a^{(1)}\\
\phi^{(1)}\\
p_{\phi}^{(1)} \end{array} \right).
\end{equation}
The normal part of the above system, after eliminating the momentum variation
$p_{\phi}^{(1)}$, and writing $x$ for $\phi^{(1)}$, is
\begin{equation}    
    q \ddot{x} + 2\dot{q}\dot{x} + 2m^2q^3x = 0,
\end{equation}
which we further simplify like before by taking $z=q$ as the new independent
variable, and using the energy condition to get
\begin{equation}
    z(E-kz^2)x''+(2E-3kz^2)x'+m^2z^3 x = 0. \label{RatVarL0}
\end{equation}
We check the physical hypersurface of $E=0$. This requires
$k\ne 0$ for otherwise the special solution would become an equilibrium point.
Introducing a new pair of variables
\begin{equation}
    w(s)=w\left(2\frac{m}{\sqrt{k}}z\right)=z^{3/2}x(z),
\end{equation}
we finally get
\begin{equation}
    \frac{\ud^2w}{\ud s^2}
    = \left(\frac14 - \frac{\kappa}{s} + \frac{4\mu^2-1}{4s^2}\right)w,
\end{equation}
with $\mu = \pm 1$, and $\kappa=0$. This is the Whittaker equation, and its
solutions are Liouvillian if, and only if,
$\left(\kappa+\mu-\frac12,\kappa-\mu-\frac12\right)$ are integers, one of them being
positive and the other negative \cite{Morales:99::}. As this is not the case here,
this finishes the proof for $k\ne0$. We recall, that because of the irregular
singular point $s=\infty$, this rules out only the rational first integrals.

Non-integrability on one energy level means no global integrability, for
the existence of another integral for all values of $E$ would imply its
existence on $E=0$. However, there might exist additional integrals for only
some, special values of the energy. It is straightforward to check with the use
of Kovacic's algorithm \cite{Kovacic:86::}, that this is not true here.
For our equation, in cases 1 and 2 of the algorithm, there is no appropriate
integer degree of a polynomial needed for the construction of the
solution, and case 3 cannot hold, because of the orders of the singular points
of the equation.

If $k=0$, a change of the dependent variable to $w(z)=zx(z)$, reduces 
equation (\ref{RatVarL0}) to
\begin{equation}
    Ew''+m^2z^2w=0,
\end{equation}
which is known not to possess Liouvillian solutions \cite{Kovacic:86::}.

We notice that when $\Lambda=E=k=0$, the system can be reduced to a two-dimensional
one. In fact, the reduction is still possible when $\Lambda\ne 0$, so we choose to
present in the next section.

%
%

\subsection{$\Lambda\ne 0$ case}

We use the nonzero constant $\Lambda$ to rescale the system as follows
\begin{equation}
\begin{aligned}
    a &= \frac{q_1}{\sqrt{\Lambda}} ,& p_a &= \frac{p_1}{\sqrt{\Lambda}},\\
    \phi & = q_2,& p_{\phi} &= \frac{p_2}{\Lambda},
\end{aligned} \label{Scaling}
\end{equation}
so that the equations become\begin{equation}
\begin{aligned}
    \dot{q}_1 &= -p_1, &&&
    \dot{p}_1 &= 2k q_1 - 4q_1^3(1 +b q_2^2) + \frac{1}{q_1^3}p_2^2,\\
    \dot{q}_2 &= \frac{1}{{q_1}^2}p_2, &&&
    \dot{p}_2 &= -2bq_2 q_1^4,
\end{aligned} \label{scaled_eq}
\end{equation}
where $b=m^2/\Lambda$. The energy integral, for the previously defined particular
solution, now reads $\mathcal{E} = E \Lambda = \frac12 \dot{q}^2 + k q^2 - q^4$,
where $q$ has been rescaled according to (\ref{Scaling}).

As before, we are interested in the variational equations, which read
\begin{equation}
\left( \begin{array}{c}
\dot{q}_1^{(1)}\\
\dot{p}_1^{(1)}\\
\dot{q}_2^{(1)}\\
\dot{p}_2^{(1)}
\end{array} \right) = \left( \begin{array}{cccc}
0 & -1 & 0 & 0 \\
2(k-6q^2) & 0 & 0 & 0 \\
0 & 0 & 0 & q^{-2} \\
0 & 0 & -2bq^4 & 0
\end{array} \right)
\left( \begin{array}{c}
q_1^{(1)}\\
p_1^{(1)}\\
q_2^{(1)}\\
p_2^{(1)} \end{array} \right),
\end{equation}
and writing $x$ for $q_2^{(1)}$, and $y$ for $p_2^{(1)}$. The normal part is
\begin{equation}
\begin{aligned}
    \dot{x} &= \frac{1}{q^2} y,\\
    \dot{y} &= -2 b q^4 x,
\end{aligned}
\end{equation}
or alternatively
\begin{equation}
\ddot{x}+2\frac{\dot{q}}{q}\dot{x}+2bq^2x = 0.
\end{equation}

\subsubsection{$\mathcal{E}=0$}

We first pick the particular solution lying on the  zero-energy level, as the global integrability
implies the integrability for this particular value of the Hamiltonian. It is
important to remember, however, that the converse is not true.

The normal variational equation is cast into a rational form by changing the
independent variable to $z=q^2/k$ (for $k\ne0$ which implies $k^2=1$), and
using the energy first integral. It then becomes a hypergeometric equation
\begin{equation}
    x'' + \frac{5z-4}{2z(z-1)}x' + \frac{b}{4z(z-1)}x=0, \label{Lambda_rat}
\end{equation}
with the respective characteristic exponents
\begin{equation}
\begin{aligned}
    z &= 0, & \rho &= -1,0\\
    z &= 1, & \rho &= 0,\frac12\\
    z &= \infty, & \rho &= \frac14(3-\sqrt{9-4b}),\frac14(3+\sqrt{9-4b}).
\end{aligned}
\end{equation}
By Kimura's theorem \cite{Kimura:69::}, the solutions of equation (\ref{Lambda_rat})
are Liouvillian if, and only if $9-4b=(2p-1)^2$, $p\in\mathbb Z$.
As before, this means that for the global integrability this condition must be
satisfied.

For $k=0$ the solution of NVE is $x_{1,2}=q^{-2\rho_{\infty 1,2}}$, and the
reduction to a two-dimensional system is possible, as mentioned before.

\subsubsection{$\mathcal E\ne0$}

The special solution, is now directly connected to the Weierstrass $\wp$
function, for if we introduce a new dependent variable $v$ with
\begin{equation}
    q^2 = \frac{1}{2} v + \frac{k}{3},
\end{equation}
the energy integral implies that it satisfies the equation
\begin{equation}
    \dot{v}^2 = 4 v^3 - g_2 v - g_3, \label{weiers_eq}
\end{equation}
where
\begin{equation}
g_2=\frac{16}{3}(k^2-3\mathcal E),\;g_3=\frac{32}{27}k(2k^2-9\mathcal E),
\end{equation}
and the discriminant $\Delta=1024 {\mathcal E}^2 (k^2 - 4\mathcal E)$, which we
take as non-zero to consider the generic case. Thus, taking
$w = q_2^{(1)} q$, and eliminating $p_2^{(1)}$ as before, the normal variational equation
reads
\begin{equation}
    \ddot{w} = [A\wp(\eta;g_2,g_3) + B]w,
\end{equation}
with $A=2-b$ and $B=-\frac{2}{3}k(1 + b)$. This is the Lam\'e differential
equation, whose Liouvillian solutions are known to fall into three mutually
exclusive cases, which are exactly those of Kovacic's algorithm:
\begin{enumerate}
\item The Lam\'e-Hermite case, with $A=n(n+1)=2-b$, $n\in\mathbb N$. This implies
that $9-4b=(2n+1)^2$. Case with $n=1$  already known to be integrable
because $b=0$ represents the massless field.

\item The Brioschi-Halphen-Crawford case, where necessarily $n$ is half an
integer, i.e. $n+\frac12=l\in\mathbb N$, and as before $9-4b=(2n+1)^2=(2l)^2$.

\item The Baldassarri case, with $n+\frac12 \in \frac13 \mathbb Z \cup
\frac14\mathbb Z\cup\frac15\mathbb Z\setminus\mathbb Z$, and
additional algebraic restrictions on $B,g_2$, and $g_3$.
\end{enumerate}
In the Lam\'e-Hermite case we have infinite number of values of  $b=2-n(n+1)$, $n\in\N$, for which the necessary conditions for the integrability given by the Morales-Ramis Theorem~\ref{thm:MR} are satisfied.  
In order to obtain stronger result we need to apply a more restrictive necessary conditions.  
Such conditions are given by higher order variational equations, see \cite{Morales:06::} for detailed exposition. Here we explain this technique on the considered problem and we will follow  \cite{Maciejewski:2005}.  

 At the beginning, it is convenient to change the variables in equations
\eqref{scaled_eq} in the following way
\begin{equation}
\begin{aligned}
    q_1 &= w_1, & p_1 &= -w_2,\\
    q_2 & = \frac{w_3}{w_1}, & p_2 &= w_1 w_4 - w_2 w_3.
\end{aligned}
\end{equation}	
Let  
\begin{equation}
 \label{eq:w}
\dot w = W(w), \quad w=(w_1,w_2,w_3,w_4), 
\end{equation} 
be the system \eqref{scaled_eq}  written in the new variables.  The advantage of new coordinates is that now 
 the variational equations split into a direct product of two Lam\'e equations
\begin{equation}
\left( \begin{array}{c}
\dot{w}_1^{(1)}\\
\dot{w}_2^{(1)}\\
\dot{w}_3^{(1)}\\
\dot{w}_4^{(1)}
\end{array} \right) = \left( \begin{array}{cccc}
0 & 1 & 0 & 0 \\
A_1\wp(\eta) + B_1 & 0 & 0 & 0 \\
0 & 0 & 0 & 1 \\
0 & 0 & A_2\wp(\eta) + B_2 & 0
\end{array} \right)
\left( \begin{array}{c}
w_1^{(1)}\\
w_2^{(1)}\\
w_3^{(1)}\\
w_4^{(1)} \end{array} \right),
\label{double_L}
\end{equation}
where $\wp(\eta)$ is the one given by equation (\ref{weiers_eq}), and
\begin{equation}
\begin{aligned}
    A_1 &= 6, & B_1 &= 2k,\\
    A_2 &= n(n+1), & B_2 &= \frac23 k(n^2+n-3).
\end{aligned}
\end{equation}
 To derive the higher order  VE's  we substitute into equation \eqref{eq:w} the
infinite formal series
\begin{equation}
    w = \varphi(\eta) +
    \epsilon w^{(1)}+\epsilon^2 w^{(2)}+\epsilon^3 w^{(3)}+\cdots,
\end{equation}
where $\varphi$ is the particular solution, and get
\begin{equation}
    \dot{w}^{(j)} =
     W'( \varphi(\eta))
    w^{(j)}+f_j(w^{(1)},\ldots,w^{(j-1)}),\qquad j=1,2,\ldots,
    \label{hve_gen}
\end{equation}
where  $W'( \varphi(\eta))$ is the matrix of right hand sides in~\eqref{double_L}, and $f_j(w^{(1)},\ldots,w^{(j-1)}) $  are vectors obtained from the Taylor expansions of components of $W(w)$. In particular we have
\begin{equation}
\begin{aligned}
    f_1 &= 0,\\
    f_2 &= \frac12 W''( \varphi(\eta))(w^{(1)} ,w^{(1)}),\\
    f_3 &= \frac16 W'''( \varphi(\eta))(w^{(1)} ,w^{(1)} ,w^{(1)})+
  W''( \varphi(\eta))( w^{(2)} ,w^{(1)}),
\end{aligned}
\end{equation}
and so on. For $j=1$ equation \eqref{double_L} is recovered. Although  $w^{(1)},\ldots,w^{(j-1)}$ enter polynomially in the right hand sides of $j$-th variational equations~\eqref{hve_gen}, there exists an appropriate framework to define their differential Galois group. In \cite{Morales:06::} it was proved that if the system is integrable, then the identity component $G_j^\circ$ of the differential Galois group $G_j$ of $j$-th variational equations is Abelian. Generally it is very difficult to determine $G_j$ for $j>1$. However, in a case when the first variational equations are a product of two Lam\'e equations having infinite differential Galois group we have an effective method to decide whether $G_j^0$ is Abelian. Namely, if a logarithmic therm appears in local solution around $\eta=0$ of $j$-th variational equations, then $G_j^0$ is not Abelian, see \cite{Morales:06::,Morales:07::} for details. 

The calculations proceed as follows. The solution of \eqref{hve_gen} is given by
\begin{equation}
    w^{(j)} = X \int X^{-1}f_j\ud\eta,
    \label{integrand}
\end{equation}
where  $f_j=f_j(w^{(1)},\ldots,w^{(j-1)})$ and  $X$ is the fundamental matrix of the homogeneous system (i.e. the first order
VE \eqref{double_L}), so that
\begin{equation}
    \dot{X} = W'(\varphi(t)) X,\qquad \det{X}\neq0.
\end{equation}
We took
\begin{equation}
    X = \left( \begin{array}{cccc}
    v_1 & v_2 & 0 & 0 \\
    \dot{v}_1 & \dot{v}_2 & 0 & 0\\
    0 & 0 & v_3 & v_4 \\
    0 & 0 & \dot{v}_3 & \dot{v}_4
    \end{array} \right),
\end{equation}
with
\begin{equation}
\begin{aligned}
    v_1 &= \eta^3 + \frac{k}{7}\eta^5 +\cdots, &
    v_2 &= -\frac{1}{5\eta^2} +\frac{k}{15} +\cdots,\\
    v_3 &= \eta^{n+1} + \frac{k(n^2+n-3)}{6n+9}\eta^{n+3} +\cdots, &
    v_4 &= -\frac{1}{(2n+1)\eta^n}+
    \frac{k(n^2+n-3)}{(2n+1)(6n-3)\eta^{n-2}}+\cdots.
\end{aligned}
\end{equation}
Next, we take as the solution of the first order  VE
\begin{equation}
    w^{(1)} = (0,0,v_4,\dot{v}_4).
\end{equation}
Then we fix $n=2$ and solve the second order VE and  we obtain the integrand of
\eqref{integrand}
for $j=3$ to be
\begin{equation}
    X^{-1}f_3 =
    (0,0,
    \frac{54}{625\eta^8}-\frac{44k}{625\eta^6}+\cdots,
    \frac{54}{125\eta^3}-\frac{128k}{875\eta}+\cdots),
\end{equation}
which produces a logarithm in $w^{(3)}$. If $k=0$, one has to find solutions of fourth order VE
 to get
\begin{equation}
    X^{-1}f_5 =
    (0,0,
    -\frac{3618}{109375\eta^{10}}-\frac{1272\mathcal E}{21875\eta^6}+\cdots,
    -\frac{3618}{21875\eta^5}-\frac{1536\mathcal E}{21875\eta}+\cdots),
\end{equation}
which proves the non-integrability, since we  assumed $\mathcal E\neq0$. 

This behaviour does not change as we increase  $n$, although it
was checked only for 10 consecutive values. We thus {\it conjecture} that for $b=2-n(n+1)$ with integer $n>1$ 
the system is not integrable. 
The procedure described is correct under assumption that the differential Galois group of the normal variational equations is not finite. We discuss this point in Appendix C, and justify that except countable many values of energy the group is not finite.

In the Brioschi case, there is another additional condition for the
integrability:  the so-called Brioschi determinant $Q_l$ is zero
\cite{Morales:99::}. Unfortunately, there is no closed formula for $Q_l$ for
general $l$, but analysing the first few values we notice a pattern:
\begin{equation}
\begin{aligned}
    Q_1 &= -\frac32 k,\\
    Q_2 &= -\frac{3}{4}(5k^2-16\mathcal E),\\
    Q_3 &= -\frac{9}{8}k(35k^2-192\mathcal E),\\
    Q_4 &= -\frac{5}{16}(2835k^4-21600k^2\mathcal E-48384\mathcal E^2),\\
    Q_5 &= -\frac{4725}{32}k(231k^4-2240k^2\mathcal E - 16384\mathcal E^2),\\
    Q_6 &= -\frac{8505}{64}(15015k^6-176400k^4\mathcal E -2802432k^2\mathcal E^2
    - 1126400\mathcal E^3).
\end{aligned}
\end{equation}
When $k=0$, $Q_l$ is zero for odd $l$, and proportional to energy, which is not
zero, for even $l$. When $k\neq0$, so that $k^2=1$, each $Q_l$ is a polynomial
in $\mathcal E$, and that gives at most a finite number of energy values for
which $Q_l=0$ and the system is potentially integrable. We, again, {\it
conjecture} that if the system is integrable (with this subsection's
assumptions) and $k=0$, then necessarily $n+\frac12$ is odd, and that if
$k^2=1$, then it is not integrable on a generic energy level.

The Baldassarri case can also be studied in more detail by means of the modular
function
\begin{equation}
    j = \frac{g_2^3}{g_2^3-27g_3^2} = 
    \frac{4(k^2-3\mathcal E)^3}{27(k^2-4\mathcal E)\mathcal E^2}=
\begin{cases}
  \frac{4(1-3\mathcal E)^3}{27(1-4\mathcal E)\mathcal E^2}&\text{for\ }k^2=1,\\
1&\text{for\ }k=0.
\end{cases}
\end{equation}
A theorem by Dwork \cite{Morales:99::} states that the number of pairs $(j,B)$
is at most finite in integrable cases. Since $j$ depends on the energy for
non-zero $k$, and $B$ depends on $m^2$, it means that a generic energy
level is not integrable for a given value of $m^2$.

\subsubsection{$\mathcal E=k=0$}
As mentioned in Section~\ref{sec:minimallyL0} in this case  we can transform the
system to a two-dimensional one.
In order to do that, time needs to be changed from the conformal to the
cosmological one $\ud\eta \rightarrow \ud t = a\ud\eta$ in the original
equations (\ref{MainEq}). We then take as the
new momenta the Hubble's function $a$  and the derivative of $\phi$
\begin{equation}
\begin{aligned}
    h &:= \frac{1}{a}\frac{\ud a}{\ud t} = -\frac{p_a}{a^2},\\
    \omega &:= \frac{\ud\phi}{\ud t} = \frac{p_{\phi}}{a^3}.
\end{aligned}
\end{equation}
(This $\omega$ is not to be confused with the one introduced in Section~2.)
Accordingly we have
\begin{equation}
\begin{aligned}
    \frac{\ud a}{\ud t} &= ah, \\
    \frac{\ud\phi}{\ud t} &= \omega, \\
    \frac{\ud h}{\ud t} &= 4\Lambda + 4m^2\phi^2-\omega^2-2h^2, \\
    \frac{\ud\omega}{\ud t} &= -2m^2\phi-3\omega h.
\end{aligned}
\end{equation}
Thus, we are left with a dynamical system in the $(h,\phi,\omega)$ space, as $a$
decouples. Furthermore, the energy integral is now
\begin{equation}
    0 = \frac12 a^4(2\Lambda + \omega^2 + 2m^2\phi^2 - h^2),
\end{equation}
so for $a(t)$ which is not trivially zero, it gives a first integral on the
reduced space. Choosing an appropriate variable $\alpha$, suggested by the
form of this integral
\begin{equation}
\begin{aligned}
    \phi &= \frac{\sqrt{h^2-2\Lambda}}{\sqrt{2}m}\sin(\alpha),\\
    \omega &= \sqrt{h^2-2\Lambda}\cos(\alpha),
\end{aligned}
\end{equation}
we finally obtain
\begin{equation}
\begin{aligned}
    \frac{\ud\alpha}{\ud t} &= \sqrt{2}\,m+3h\sin(\alpha)\cos(\alpha),\\
    \frac{\ud h}{\ud t} &= -3(h^2-2\Lambda)\cos^2(\alpha).
\end{aligned}
\end{equation}

The problem of such reduction was
also discussed in \cite{Faraoni:2006sr}. It is argued that there can be no
chaos in this system, but its integrability -- which would be one more first
integral -- remains unresolved.

\section{Analysis of the conformally coupled field}
\label{sec:conformally}

\subsection{Known integrable families}

There are four known cases when the system has an additional first integral,
functionally independent of the Hamiltonian. They were found by applying the so-called ARS
algorithm basing on the Painlev\'e analysis \cite{Ablowitz:80::a}. Table~\ref{tab:conformally}  summarises those results.
\begin{center}
\begin{table}
\begin{center}
\begin{tabular}{|c|c|c|c|}
\hline
 case &$k$&$\Lambda$&$m^2$\\\hline
\rm{(1)}&$0,\pm 1$&$\Lambda=\lambda$&$m^2=-3\Lambda$\\\hline
\rm{(2)}&$0,\pm 1$&$\Lambda=\lambda$&$m^2=-\Lambda$\\\hline
\rm{(3)}&$0$&$\Lambda=16 \lambda$&$m^2=-6\lambda$\\\hline
\rm{(4)}&$0$&$\Lambda=8 \lambda$&$m^2=-3\lambda$\\\hline
\end{tabular}
\end{center}
\caption{Known integrable cases for the conformally coupled field}
\label{tab:conformally}
\end{table}
\end{center}
And the respective integrals of the systems are
\begin{equation}
\begin{split}
(1)&\begin{cases}
H=\dfrac{1}{2}(p_2^2-p_1^2)+\dfrac{k}{2}(q_2^2-q_1^2)-
\dfrac{m^2}{12}(q_1^4-6q_1^2q_2^2+q_2^4),\\
I=p_1p_2+\dfrac{1}{3}(m^2(q_2^2-q_1^2)-3k),
\end{cases}\\
(2)&\begin{cases}
H=\dfrac{1}{2}(p_2^2-p_1^2)+\dfrac{k}{2}(q_2^2-q_1^2)-
\dfrac{m^2}{4}(q_2^2-q_1^2)^2,\\
I=q_1p_2+q_2p_1,
\end{cases}\\
(3)&\begin{cases}
H=\dfrac{1}{2}(p_2^2-p_1^2)-
\dfrac{m^2}{24}(16q_1^4-12q_1^2q_2^2+q_2^4),\\
I=(q_1p_2+q_2p_1)p_2+\dfrac{m^2}{6}q_1q_2^2(q_2^2-2q_1^2),
\end{cases}\\
(4)&\begin{cases}
H=\dfrac{1}{2}(p_2^2-p_1^2)-
\dfrac{m^2}{12}(8q_1^4-6q_1^2q_2^2+q_2^4),\\
I=p_2^4+\dfrac{m^2q_2^2}{3}\left[4q_1q_2p_1p_2+q_2^2p_1^2-(q_2^2-6q_1^2)p_2^2+
\dfrac{1}{12}q_2^2(q_2^2-2q_1^2)^2\right].
\end{cases}
\end{split}
\end{equation}

In this work, we will show, that the above are the only integrable cases, when
$m\ne0$. An important point to note is that there is a complete symmetry with
respect to interchanging $\Lambda$ and $\lambda$. It is a consequence of the
fact, that there exists a canonical transformation of the form
\begin{equation}
\begin{aligned}
    p_1 \rightarrow i\,p_1,& \quad q_1 \rightarrow -i\,q_1,\\
    p_2\rightarrow p_2,& \quad q_2 \rightarrow q_2,
\end{aligned}
\end{equation}
that changes the Hamiltonian into
\begin{equation}
    H = \frac12\left(p_1^2 + p_2^2\right) + \frac12\left[k(q_1^2 + q_2^2) 
    - m^2 q_1^2q_2^2\right] +\frac14\left(\Lambda q_1^4 +\lambda q_2^4\right),
\end{equation}
which is the same after swapping the indices. We shall use this form of $H$,
where the kinetic part is in the natural form, to make the use of some already
existing theorems more straightforward.

\subsection{Integrability of the reduced problem}

It is possible to give stringent conditions for integrability of the system, by
considering a reduced Hamiltonian. Namely, we can separate potential V into
homogeneous parts of degree 2 and 4:
\begin{equation}
\begin{aligned}
    V&=V_{h2}+V_{h4},\\
    V_{h2}&=\frac12 k\left(q_1^2+q_2^2\right),\\
    V_{h4}&=\frac14\left( -2m^2 q_1^2q_2^2 + \Lambda q_1^2 +
    \lambda q_2^4\right).
\end{aligned} \label{V_parts}
\end{equation}
The following fact is crucial in our considerations: if a potential $V$ is
integrable then its highest order as well as the lowest order parts are also
integrable. This fact needs some additional justification as its several known
proofs are not correct. In fact, consider potential
$V=V_{\mathrm{min}}+\cdots+V_{\mathrm{max}}$, where   $V_{\mathrm{min}}$ and
$V_{\mathrm{max}}$ are homogeneous parts of $V=V(q)$, $q\in\mathbb{C}^n$ of the
lowest and the highest degree, respectively. Assume that it admits meromorphic
commuting independent first integrals $F_1,\ldots, F_n$.  If $F_i=R_i/S_i$ for
certain holomorphic functions $R_i$ and $S_i$,  then we set $f_i =r_i/s_i$,
where $r_i$ and $s_i$ are the lowest order terms of expansions of $R_i$ and
$S_i$ into the power series.  It is easy to show that $f_i$ are first integrals
of $V_{\mathrm{min}}$. However, we cannot claim  that they are functionally
independent.  Fortunately we can  use in the described situation the Ziglin
Lemma~\cite{Ziglin:82::b} which guarantees that we can always choose first
integrals $F_i$ in such a way that their lading terms $f_i$ are functionally
independent. A more complicated situation arises with the integrability of
$V_{\mathrm{max}}$. Here we have to assume that $V$ is integrable with rational
first integrals in order to distinguish their highest order terms. Then
we need also an appropriate version of the Ziglin~Lemma. Proofs of these facts
will be published elsewhere.  

In
our case if $V$ given by (\ref{V_parts}) is integrable then $V_{h2}$ and
$V_{h4}$ must also be
integrable. $V_{h2}$ is the potential of the two-dimensional harmonic
oscillator, thus, it is trivially integrable. However, the homogeneous part
$V_{h4}$ gives strong integrability restrictions for the whole
potential V. We will call $V_{h4}$ the reduced potential and denote it by $\widehat V$.

Thus we effectively set $k=0$, and are now in position to exercise known
theorems concerning homogeneous potentials depending on  two variables. In particular the
complete analysis for degree 4 has been completed in \cite{Maciejewski:2005}.

In order to identify our potential with some of the list given in that paper,
we have to check how many Darboux points there exist, and what are the values
of parameters $\Lambda$, $\lambda$ and $m$ that give potentials equivalent to 
particular families.

We say that a non-zero point $(q_1,q_2)=\boldsymbol d$ is a Darboux point of
the potential $\widehat V(q_1,q_2)$ when it satisfies the equation
\begin{equation}
    \widehat V'(\boldsymbol d) = \gamma\boldsymbol d,
\end{equation}
where $\gamma\in\mathbb C^{\ast}=\mathbb C \setminus\{0\}$. Such a point
corresponds to a particular solution of the form
\begin{equation}
    \boldsymbol q(\eta) = f(\eta)\boldsymbol d,\quad
    \boldsymbol p(\eta) = \dot f(\eta)\boldsymbol d,
\end{equation}
with $f(\eta)$ satisfying a differential equation that for a potential of degree 4  takes the form
\begin{equation}
    \ddot f(\eta)=-\gamma f(\eta)^{3}.
\end{equation}

As explained in Section~3, particular solutions allow for
studying the variational equations along them, and yield necessary conditions
for existence of additional first integrals. However, the major simplification
discovered in \cite{Maciejewski:2005} is that additionally there is only a finite number
of parameters' sets (or non-equivalent potentials) corresponding to integrable
cases.

Following the cited paper's exposition and notation, we take $\mathcal I_{4,2}$
and $\mathcal I_{4,3}$ to be the sets of integrable homogeneous potentials of
degree 4 with 2 and 3 simple Darboux points respectively. We recall also four
characteristic potentials thereof
\begin{equation}
\begin{aligned}
    V_3 &= \frac14aq_1^4+\frac13bq_1^3q_2+\frac14(q_1^2+q_2^2)^2,\\
    V_4 &= \frac14aq_1^4+q_2^4,\\
    V_5 &= 4q_1^4+3q_1^2q_2^2+\frac14q_2^4,\\
    V_6 &= 2q_1^4+\frac32q_1^2q_2^2+\frac14q_2^4,
\end{aligned}
\end{equation}
where $a$ and $b$ denote (for the sake of this paragraph) arbitrary complex
numbers.

We find that our potential has:
\begin{enumerate}
\item{Four simple Darboux points,} when
$\Lambda(m^2+\Lambda)(m^2+\lambda)\ne0$, and
$\Lambda\lambda\ne m^4$. The only integrable cases are:
\begin{enumerate}
\item{$\lambda=\Lambda=-\frac13m^2$} ($\widehat V$ is equivalent to $V_4$),
\item{$\lambda=-\frac83m^2,\quad \Lambda=-\frac16m^2$} ($\widehat V$ is equivalent to
$V_5$),
\item{$\lambda=-\frac83m^2,\quad \Lambda=-\frac13m^2$} ($\widehat V$ is equivalent to
$V_6$).
\end{enumerate}
\item{Three simple Darboux points,} when $\Lambda=0$, and
$\lambda(m^2+\lambda)\ne0$. There are no integrable families here as
$\mathcal I_{4,3} =\emptyset$.
\item{Two simple Darboux points,} when either $\Lambda=\frac{m^4}{\lambda}$ and
$\lambda(m^2+\lambda)\ne0$, or $\Lambda=\lambda=0$. Again, no integrable
families are present here because $\mathcal I_{4,2}=\emptyset$.
\item{A triple Darboux point,} when $\Lambda=-m^2$. Additionally there is a
simple Darboux point when $\lambda\ne0$. The potential is equivalent to $V_3$
and is only integrable when $\lambda=-m^2$.
\end{enumerate}

There are two immediate implications that follow. Firstly, that the main system
itself with $k=0$ is only integrable in those four cases, and the respective
first integrals are known, as given in the table. Secondly,
as it was shown in \cite{Hietarinta}
those cases are the only ones which could be integrable
when $k\ne0$. This happens because the integrability of the full potential
implies the integrability of the homogeneous parts of the maximal and minimal
degree (the latter is trivially solvable in our case).

As the table shows, when the potential is equivalent to $V_3$
(or, to be precise, its integrable subcase) or $V_4$, the second first integral
is known; but $V_5$ and $V_6$ only have known first integrals with zero curvature.
And as was shown in \cite{Boucher}, for $k=1$, the values of $\Lambda$ and
$\lambda$ are those of $V_5$ or $V_6$ forbid integrability. This is easily
extended to the $k=-1$ case, since after the change of variables
\begin{equation}
    q_j \rightarrow e^{i\pi/4}q_j,\quad p_j \rightarrow e^{-i\pi/4}p_j,\quad j=1,2,
\end{equation}
we obtain a system with the sign of $k$ changed, but the ratios $m^2/\Lambda$
and $m^2/\lambda$ the same. Thus, concerning the conjecture of the quoted
paper, our results for $k\ne0$ enable us to state, that it is true, when the
rational integrability is considered.

However, the above considerations assume that the energy value is generic, so
that the particular solution is a non-degenerate elliptic function. As stressed
before, this does not preclude the existence of an additional first
integral on the physically crucial zero-energy level.

\subsection{Integrability on the zero-energy level}

We choose not to investigate the Darboux points, but the variational equations
directly, as they are considerably simpler in this case.
The Hamiltonian equations of (\ref{ham_orig1}) are
\begin{equation}
\begin{aligned}
\dot q_1&=p_1,&\quad \dot p_1&=-kq_1+m^2q_1q_2^2-\Lambda q_1^3,\\
\dot q_2&=p_2,&\quad \dot p_2&=-kq_2+m^2q_1^2q_2-\lambda q_2^3,
\end{aligned}
\end{equation}
and they admit three invariant planes as was shown in
\cite{Maciejewski:00::b}. They are
\begin{equation}
\begin{split}
\Pi_k&=\{(q_1,q_2,p_1,p_2)\in\mathbb{C}^4\,|\,q_k=0\wedge p_k=0\},\qquad
k=1,2,\\
\Pi_3&=\{(q_1,q_2,p_1,p_2)\in\mathbb{C}^4\,|\,q_2=\alpha q_1\wedge p_2=-\alpha
p_1\},\qquad \alpha^2=\dfrac{m^2+\Lambda}{m^2+\lambda}.
\end{split}
\end{equation}
Obviously two particular solutions are
\begin{equation}
\begin{aligned}
&\{q_1=p_1=0,\,q_2=q_2(\eta),\,p_2=q_2'(\eta)\},&\quad
0&=\frac{1}{2}\left(p_2^2+kq_2^2+\frac{\lambda}{2}q_2^4\right),\\
&\{q_2=p_2=0,\,q_1=q_1(\eta),\,p_1=q_1'(\eta)\},&\quad
0&=\frac{1}{2}\left(+p_1^2+kq_1^2+\frac{\Lambda}{2}q_1^4\right),
\end{aligned}
\end{equation}
and in order to find the third particular solution we make a canonical change
of variables
\begin{equation}
(q_1,q_2,p_1,p_2)^T=\mathbb{B}(Q_1,Q_2,P_1,P_2)^T,
\end{equation}
where symplectic matrix $\mathbb{B}$ has the block structure
\begin{equation}
\mathbb{B}=\begin{pmatrix}
\mathbb{A}&\mathbb{O}\\
\mathbb{O}&\mathbb{A}^T
\end{pmatrix},\qquad
\mathbb{A}=\begin{pmatrix}
-b&-a\\
-a&b
\end{pmatrix},\qquad
\mathbb{O}=\begin{pmatrix}
0&0\\
0&0
\end{pmatrix},
\end{equation}
and $a$ and $b$ are defined by
\begin{equation}
a=\sqrt{\dfrac{m^2+\Lambda}{2m^2+\lambda+\Lambda}},\qquad
b=\sqrt{\dfrac{m^2+\lambda}{2m^2+\lambda+\Lambda}}.
\end{equation}
Let us introduce five quantities
\begin{equation}
\begin{split}
\alpha_1&= 2m^2+\lambda+\Lambda,\quad
\alpha_2=3\lambda\Lambda+2m^2(\lambda+\Lambda)+m^4,\quad
\alpha_3=\sqrt{(\lambda+m^2)(\Lambda+m^2)},\\
\alpha_4&=\lambda^2+\Lambda^2-\lambda\Lambda-m^4,\qquad
\alpha_5=\lambda\Lambda-m^4.
\end{split}
\end{equation}
Then, in the new variables, Hamiltonian (\ref{ham_orig1}) has the form
\begin{equation}
H=\frac{1}{2}\left[P_1^2+P_2^2+k(Q_1^2+Q_2^2)\right]+
\frac{1}{4\alpha_1}
\left[\alpha_5Q_1^4+2\alpha_2
Q_1^2Q_2^2+4(\Lambda-\lambda)\alpha_3Q_1Q_2^3
+\alpha_4Q_2^4\right],
\end{equation}
and the Hamiltonian equations read
\begin{equation}
\begin{aligned}
\dot Q_1&=P_1,&\quad
\dot P_1&=-kQ_1-\frac{1}{\alpha_1}\left[\alpha_5Q_1^3
+\alpha_2Q_1Q_2^2+(\Lambda-\lambda)\alpha_3Q_2^3\right],\\
\dot Q_2&=P_2,&\quad
\dot P_2&=-
kQ_2-\frac{1}{\alpha_1}\left[\alpha_2Q_1^2Q_2+3(\Lambda-\lambda)\alpha_3
Q_1Q_2^2+\alpha_4 Q_2^3\right].
\end{aligned}
\end{equation}
Thus, the third particular solution can be seen to be
\begin{equation}
    \{Q_2=P_2=0,\,Q_1=Q_1(\eta),\,P_1=Q_1'(\eta)\},\qquad
    0=\frac{1}{2}\left(P_1^2+kQ_1^2+\dfrac{\alpha_5}{2\alpha_1}Q_1^4\right).
\end{equation}
Of course, this is only valid for $\alpha_1\ne0$. We investigate what happens
when $\lambda+\Lambda=-2m^2$ at the end of this section.

Normal variational equations (NVE's) along those three solutions (in the position
variables) are
\begin{equation}
\begin{aligned}
    \xi''(\eta) &= \left[-k+m^2 q(\eta)^2\right]\xi(\eta),\\
    \xi''(\eta) &= \left[-k+m^2 q(\eta)^2\right]\xi(\eta),\\
    \xi''(\eta) &= \left[-k - \frac{\alpha_2}{\alpha_1}q(\eta)^2\right]\xi(\eta),
\end{aligned}
\end{equation}
where $q(\eta)$ is one of $\{q_1(\eta),\,q_2(\eta),\,Q_1(\eta)\}$, depending on the
respective particular solution.

We will consider the $k=0$ case first. Changing the independent variable to
$z=q(\eta)^2$, all the NVE's are reduced to the following
\begin{equation}
    2z^2\xi''(z) + 3z\,\xi'(z) - \lambda_i\,\xi(z) = 0,
\end{equation}
whose solution is
\begin{equation}
    \xi(z) = z^{-(1\pm\sqrt{1+8\lambda_i})/4},
\end{equation}
where we have introduced three important quantities
\begin{equation}
    \lambda_1 = -\frac{m^2}{\Lambda},\qquad
    \lambda_2 = -\frac{m^2}{\lambda},\qquad
    \lambda_3 = \frac{\alpha_2}{\alpha_5} = 
    \frac{3-2(\lambda_1+\lambda_2)+\lambda_1\lambda2}{1-\lambda_1\lambda_2}.
    \label{lamb_def}
\end{equation}

Note, that if any of $\Lambda$, $\lambda$ or $\alpha_5$ is zero, the
corresponding particular solution is constant and cannot be used to restrict
the problem's integrability. Thus, we are left with the $E=k=0$ case as
suspected to be integrable.

When we assume $k\ne0$, or more precisely $k^2=1$, and introduce the same
independent variable $z$ as before, the NVE's read
\begin{equation}
\begin{aligned}
   & 2z^2(\Lambda z+2k)\xi''(z) + z(3\Lambda z+4k)\xi'(z) + (m^2z-k)\xi(z)=0,\\
  &  2z^2(\lambda z+2k)\xi''(z) + z(3\lambda z+4k)\xi'(z) + (m^2z-k)\xi(z)=0,\\
  &  2z^2\left(\frac{\alpha_5}{\alpha_1}z+2k\right)\xi''(z)
    + z\left(3\frac{\alpha_5}{\alpha_1} z+4k\right)\xi'(z)
   - \left(\frac{\alpha_2}{\alpha_1}z+k\right)\xi(z) = 0.
\end{aligned}
\end{equation}
First, let us observe that unlike in the previous case, when any of $\Lambda$,
$\lambda$ or $\alpha_5$ is zero, the system is not integrable. This happens,
because  then the NVE's  becomes the Bessel equation 
\begin{equation}
    s^2\xi''(s)+s\xi'(s)+(s^2-n^2)\xi(s)=0,
\end{equation}
with $n=1$ and in a new variable $s=m\sqrt{z/k}$ (for the first two equations) or
$s=m\sqrt{-2z/k}$ (for the third equation). The Bessel equation is known not
to possess Liouvillian solutions for $n=1$ \cite{Kovacic:86::}. Together with
the results of the previous section this leads us to the following lemma.
\begin{lemma}
System (\ref{ham_orig1}) considered on the zero or generic
energy level with $k^2=1$ is not integrable when
$\Lambda$ or $\lambda$ is zero. Additionally for $\lambda+\Lambda\ne-2m^2$, it is not
integrable when $\lambda\Lambda=m^4$.
\end{lemma}

Assuming that none of those constants is zero, we rescale the variable $z$ in
the three equations with
\begin{equation}
    z \rightarrow -\frac{2k}{\Lambda}z,\quad z\rightarrow-\frac{2k}{\lambda}z,
    \quad z\rightarrow \frac{2k\alpha_1}{\alpha_5}z,
\end{equation}
respectively, so that all three are transformed into a Riemann P equation of
the form
\begin{equation}
    \xi''(z) +
    \left(\frac{1-\delta-\delta'}{z}+\frac{1-\gamma-\gamma'}{z-1}\right)\xi'(z)
    +\left[\frac{\delta\delta'}{z^2}+\frac{\gamma\gamma'}{(z-1)^2}+
    \frac{\beta\beta'-\delta\delta'-\gamma\gamma'}{z(z-1)}\right]\xi(z) = 0,
\end{equation}
with the following pairs of exponents $(\delta,\delta')$, $(\gamma,\gamma')$,
$(\beta,\beta')$ at its singular points
\begin{equation}
    \left(\frac12,-\frac12\right),\quad
    \left(\frac12,0\right),\quad
    \left(\frac14+\frac14\sqrt{1+8\lambda_i},
    \frac14-\frac14\sqrt{1+8\lambda_i}\right),\qquad i=1,2,3.
\end{equation}
Using Kimura's results on solvability of the Riemann P equation \cite{Kimura:69::}
we check when the difference of the
exponents give us cases with the necessary conditions for integrability 
satisfied, and find that the parameters must belong to
the following families
\begin{equation}
    \lambda_i = \frac{l_i(l_i+1)}{2},\quad l_i\in\mathbb Z,\quad i=1,2,3.
\label{e0k1con}
\end{equation}
These polynomials in $l_i$ are invariant with respect to the change
$l\rightarrow-l-1$, so it is enough to consider non-negative values only.
Furthermore, $\lambda_1$ and $\lambda_2$ cannot be equal to zero, as $m^2\ne0$,
so $l_1$ and $l_2$ need to be strictly positive.

This result can be refined still. First, let us notice, that $\lambda_i$ are
not independent and the relation between them is
\begin{equation}
    \frac{1}{\lambda_1-1}+\frac{1}{\lambda_2-1}+
    \frac{2}{\lambda_3-1} = -1, \label{relation0}
\end{equation}
provided $\alpha_1\ne0$ and $\alpha_5\ne0$.
In the above form we had to exclude the possibility of $\lambda_i=1$, so we
consider it separately.

Both of $\lambda_1$ and $\lambda_2$ cannot  be
equal to 1, as that would mean $\alpha_5=0$ and we have shown that then
the equations are non-integrable if additionally $\alpha_1\ne0$. The
$\alpha_1=0$ case is described below.

If only one of $\lambda_i$, say
$\lambda_1$ is 1, then necessarily $\lambda_3=1$, which follows from the
definition (\ref{lamb_def}), and the only possibly
integrable cases are those with $\lambda_2$ satisfying (\ref{relation0}) with
$l_2\geq2$. The same holds when $\lambda_1$ and $\lambda_2$ are interchanged.
Also, $\lambda_3=1$ requires that one of the remaining $\lambda_i$ is 1.

When $l_1$ and $l_2$ are taken to be grater than 1, $\lambda_1$ and $\lambda_2$
are positive, so the relation (\ref{relation0}) requires that $2/(\lambda_3-1)$
is negative. This only happens for $l_3=0$ and it follows that $l_1=l_2=2$,
which is exactly the first known integrable case.
Since $1/(\lambda_1-1)$ and $1/(\lambda_2-1)$ are positive and tend to zero
monotonically as $l_i\geq2$ tends to infinity, there are no other solutions,
and no other integrable sets of parameters.

Finally, we turn to see what happens when $\alpha_1=0$, i.e.
$\Lambda+\lambda=-2m^2$. This is equivalent to
\begin{equation}
    \frac{1}{\lambda_1}+\frac{1}{\lambda_2}=2,
\end{equation}
provided $\lambda\ne0$ and $\Lambda\ne0$ and the same two conditions of (\ref{e0k1con}) hold because the first two
variational equations can still be used. It is straightforward to check that
the only integer solution of
\begin{equation}
    \frac{1}{l_1(l_1+1)} + \frac{1}{l_2(l_2+1)} = 1
\end{equation}
is $l_1=l_2=1$ (so, incidentally, $\alpha_5=0$), which we recognise as the
second case of our table.

\section{Conclusions}
\label{sec:conclusions}

The main results of our paper can be summarised as follows.

For the minimally coupled scalar fields, given by the Hamiltonian
\begin{equation}
    H = \frac12 \left(-p_1^2 + \frac{1}{q_1^2}p_2^2 \right) - k q_1^2 + \Lambda q_1^4
    + m^2q_2^2 q_1^4,
\end{equation}
we have:
\begin{theorem}
For $\Lambda=0$, if the system is integrable then necessarily $E=k=0$.
\end{theorem}

\begin{theorem}
When $\Lambda\ne0$, if the system is integrable on a generic energy level then
either
\begin{enumerate}
\item $9-4m^2/\Lambda=l^2$ for some $l\in\mathbb Z$, or
\item $k=0$ and $9-4m^2/\Lambda=(2n+1)^2$ for $n+\frac12\in
\frac13 \mathbb Z \cup\frac14\mathbb Z\cup\frac15\mathbb Z\setminus\mathbb Z$.
\end{enumerate}
\label{thm:mini}
\end{theorem}

\begin{conjecture}
Suppose $\Lambda\ne0$, and let $n$ be an integer satisfying $9-4m^2/\Lambda=(2n+1)^2$. If the
system is integrable on a generic energy level $E\neq 0$, then either
\begin{enumerate}
\item $n=1$ or $n=-2$ ($m=0$ in both cases), or
\item $k=0$ and $9-4m^2/\Lambda=(2l)^2$ with $l$ an odd integer, or
\item $k=0$ and $n+\frac12\in \frac13 \mathbb Z \cup
\frac14\mathbb Z\cup\frac15\mathbb Z\setminus\mathbb Z$.
\end{enumerate}
\end{conjecture}
Note that this is more restrictive than  Theorem~\ref{thm:mini}, as
 case 1 of this theorem admits more values of $n$ than the conjecture's cases 1
and 2 put together.

\begin{theorem}
For the zero energy level, and provided that $\Lambda\ne0$, if the
system is integrable then either
\begin{enumerate}
\item $k=0$, or
\item $9-4m^2/\Lambda=(2n+1)^2$, $n\in\mathbb Z$.
\end{enumerate}
\end{theorem}

While for the conformally coupled scalar fields, given by the Hamiltonian
\begin{equation}
    H = \frac12\left(-p_1^2 + p_2^2\right) + \frac12\left[k(-q_1^2 + q_2^2) 
    + m^2 q_1^2q_2^2\right] +\frac14\left(\Lambda q_1^4 +\lambda q_2^4\right),
\end{equation}
we have:
\begin{theorem}
The system restricted to a  generic energy level  $E\neq 0$ is integrable if,
and only if,
\begin{enumerate}
\item $k=0$, and its parameters belong to the four families listed in Table~\ref{tab:conformally}. 
Otherwise there exists no additional, meromorphic integral.
\item $k^2=1$, and its parameters belong to the first two families of Table~\ref{tab:conformally}.
 Other than that, there exists no additional, rational first integral.
\end{enumerate}
\end{theorem}
The second part of the above theorem can be strengthened to meromorphic
first integrals, although not for all values of the parameters, as described in
\cite{Boucher}.

\begin{theorem}
If the system restricted to the zero energy level
is integrable, then either
\begin{enumerate}
\item $k=0$, or
\item $k^2=1$ and its parameters belong to the first two families of Table~\ref{tab:conformally},
or
\item $k^2=1$ and one of $\{\lambda_1$, $\lambda_2\}$ is equal to 1, and the
other satisfies the condition (\ref{e0k1con}) with $l_i\geq2$.
\end{enumerate}
Otherwise, the system is not meromorphically integrable. In particular this
means, that for $k^2=1$ if at least one of $\Lambda$ and $\lambda$ is zero, then the
system is non-integrable.
\end{theorem}

These are, however, only necessary and not sufficient conditions, so that the
system might still prove not to be integrable at all. In particular, the
numerical search for chaos suggests both the lack of global first integrals,
and crucial differences in the behaviour of the system for real and imaginary
values of the variables. This might be a clue, that the system might have first
integrals which are not analytic, and thus not prolongable to the complex
domain. A system with similar property was studies by the authors in
\cite{Maciejewski:2005yi}.

The immediate physical consequences of the non-integrability is the non-existence
of constants or motion (by definition) or, in other words, laws of
conservation. This results not only in the complexity of evolution but also in
a harder descriptive approach to a physical system which does not possess any
global, well defined, preserved quantities like total charge or spin (in
general -- we have not considered such quantities in the present work). It is
also obvious that direct integration, or obtaining the solutions in closed
forms by means of elementary functions is out of question with non-integrable
problems.

Of course, depending on the properties of the first integrals, we might get
quite different results, and the requirement of meromorphicity or rationality
is still very restrictive. As described in the introduction, this leaves open
the question of existence of real-analytic first integrals. Also we recall that
physically the scale factor $a$ cannot even assume negative values, and some
authors argue that when cosmological (instead of conformal) time is used, the
evolution is not, in essence, chaotic \cite{Castagnino:2001xp}. Thus, we would
like to stress that Liouvillian integrability is a mathematical property of the
system, and often the methods used to study it require the complexification of
variables. This means that when restricted to the narrower, physical domain,
the dynamics might be much simpler. And in particular we might be interested in
a particular trajectory whose behaviour is far from generic. It is no surprise
then, that the dynamics of our system when restricted to $a>0$ might appear
regular. It should still be noted that the notion of chaos, although frequently
associated with the integrability, has not yet been successfully conflated with it.
And that a regular evolution is not necessarily integrable.

\section*{Acknowledgements}
A very special thanks to the referees of this paper for  many suggestions of the improvements of this article.

For the first three authors this research was supported by
grant No. N N202 2126 33 of Ministry of Science and Higher Education of
Poland.
For the second author this research has been partially supported by the
European Community project GIFT  (NEST-Adventure Project no. 5006), by
Projet de l'Agence National de la Recherche ``Int\'egrabilit\'e r\'eelle et
complexe en m\'ecanique hamiltonienne'' N$^\circ$~JC05$_-$41465 and by the grant UMK 414-A. And for the
fourth author, by the Marie Curie Actions Transfer of Knowledge project COCOS
(contract MTKD-CT-2004-517186).

\section*{Appendix A. Massless minimal field}

When $m=0$, the Hamilton-Jacobi equation for the main Hamiltonian
(\ref{MainHam}) will become separable, because it can be written as
\begin{equation}
    Ea^2 = \frac12\left(\frac{\partial W}{\partial\phi}\right)^2
    - \frac12 a^2 \left(\frac{\partial W}{\partial a}\right)^2
    - ka^4 +\Lambda a^6 +\frac{\omega^2}{\phi^2}, \label{EqHJ}
\end{equation}
with the full generating function $S=W-E\eta$. Assuming $W=A(a)+F(\phi)$,
equation (\ref{EqHJ}) can be solved with
\begin{equation}
\begin{aligned}
    F(\phi) &= \int\sqrt{2\left(J-\frac{\omega^2}{\phi^2}\right)}\ud\phi,\\
    A(a) &= \int\sqrt{2\left(\Lambda a^4-ka^2-E+\frac{J}{a^2}\right)}\ud a,
\end{aligned}
\end{equation}
where $J$ is a constant of integration.
The first equation of motion can then be deduced from
\begin{equation}
    \frac{\partial W}{\partial E} - \eta =
    \int\frac{\ud a}{\sqrt{2\left(\Lambda a^4-ka^2-E+\frac{J}{a^2}\right)}} =
    {\rm const},
\end{equation}
which can be rewritten as
\begin{equation}
    \left(\frac{\ud a}{\ud\eta}\right)^2 =
    2\left(\Lambda a^4-ka^2-E+\frac{J}{a^2}\right).
\end{equation}
Or, introducing a new variable $v=a^2$, as
\begin{equation}
    \left(\frac{\ud v}{\ud\eta}\right)^2 =
    8(\Lambda v^3-kv^2-Ev+J), \label{EqWei}
\end{equation}
so that the general solution is
\begin{equation}
    a^2 = v = \frac{1}{2\Lambda}\wp(\eta-\eta_0;g_2,g_3)+\frac{k}{3\Lambda},
\end{equation}
where
\begin{equation}
\begin{aligned}
    g_2 &= \frac{16}{3}k^2+16\Lambda E \\
    g_3 &= \frac{32}{3}\Lambda kE+\frac{64}{27}k^3-32\Lambda^2J,
\end{aligned}
\end{equation}
and $\eta_0$ is the constant of integration.
Of course, for $\Lambda=0$, equation (\ref{EqWei}) admits solutions in terms of
circular functions.

The equation for $\phi(\eta)$ is the following
\begin{equation}
    \frac{\partial W}{\partial J} =
    \int\frac{\ud\phi}{\sqrt{2\left(J-\frac{\omega^2}{\phi^2}\right)}} +
    \int\frac{\ud a}{a^2\sqrt{2\left(\Lambda a^4-ka^2-E+\frac{J}{a^2}\right)}} =
    {\rm const}, \label{Int}
\end{equation}
which we simplify by using the just obtained solution for $v(\eta)$ to get
\begin{equation}
    {\rm const} = \frac{\sqrt{J\phi^2-\omega^2}}{\sqrt2J} + 
    \int\frac{\ud\eta}{2v}.
\end{equation}
As $v$ is an elliptic function of order two, the second integral can be
evaluated by means of the Weierstrass zeta and sigma functions to yield
\begin{equation}
    {\rm const} = \frac{\sqrt{J\phi^2-\omega^2}}{\sqrt2J} +
    \frac{1}{4\sqrt{2J}}\big[\zeta(\eta_1)-\zeta(\eta_2)\big]\eta +
    \frac{1}{4\sqrt{2J}}
    \ln\left[\frac{\sigma(\eta-\eta_1)}{\sigma(\eta-\eta_2)}\right],
\end{equation}
where $\eta_{1,2}$, are the zeroes of $v(\eta)$, given by
\begin{equation}
    3\wp(\eta_{1,2};g_2,g_3)=-2k,
\end{equation}
and the constant of integration can be determined from the boundary conditions
on the field $\phi$. The functions $\zeta$ and $\sigma$ are defined as follows
\begin{equation}
\begin{split}
    -\zeta'(z)=\wp(z),\quad\lim_{z\rightarrow 0}\left(\zeta(z)-\tfrac1z\right)=0,\\
    \frac{\sigma'(z)}{\sigma(z)}=\zeta(z),\quad\lim_{z\rightarrow 0}\frac{\sigma(z)}{z}=1.
\end{split}
\end{equation}
Again, for $J=0$, the integrals in (\ref{Int}) reduce to
simpler functions.

\section*{Appendix B. Massless conformal field}

For $m=0$ we can separately solve equations for each variable, so that we have
\begin{equation}
\begin{aligned}
    E_1 &= -\frac12\dot a^2 - \frac12ka^2 +\frac14\Lambda a^4,\\
    E_2 &= \frac12\dot\phi^2 + \frac12\frac{\omega^2}{\phi^2}+
    \frac12k\phi^2 + \frac14\lambda\phi^4,
\end{aligned}
\end{equation}
with $E_1+E_2=E$ being the total energy. The first of these is immediately
solved, when we substitute $v_1=a^2$ to get
\begin{equation}
    \dot v_1^2 = 2\Lambda v_1^3 - 4kv_1^2 -8E_1 v_1,
\end{equation}
whose solution is
\begin{equation}
    v_1(\eta) = \frac{2}{\Lambda}\wp(\eta-\eta_1;g_2,g_3) +\frac{2k}{3\Lambda},
\end{equation}
with $\eta_1$ the integration constant and
\begin{equation}
    g_2 = \frac43k^2 + 4\Lambda E_1,\quad g_3 = \frac{8}{27}k^3 +
    \frac43k\Lambda E_1.
\end{equation}
Of course, when $\Lambda=0$ the Weierstrass function $\wp$ reduces to
a trigonometric function.

Similarly, for the other variable, we substitute $v_2=\phi^2$ and obtain
\begin{equation}
    \dot v_2^2 = -2\lambda v^3 - 4kv^2 +8E_2v -4\omega^2,
\end{equation}
whose solution is
\begin{equation}
    v_2(\eta) = -\frac{2}{\lambda}\wp(\eta-\eta_2;g_4,g_5) -
    \frac{2k}{3\lambda},
\end{equation}
where
\begin{equation}
    g_4=\frac43k^2+4\lambda E_2,\quad g_5=\frac{8}{27}k^3 + \frac43k\lambda E_2
    +\lambda^2\omega^2,
\end{equation}
and $\eta_2$ is the integration constant. As before, for $\lambda=0$ the
solution degenerates to trigonometric functions.

\section*{Appendix C. Lam\'e equation in the  Lam\'e-Hermite case }
Let us consider Lam\'e equation 
\begin{equation}
 \label{eq:LH}
\frac{\rmd ^2y}{\rmd t^2}=(n(n+1)\wp(t) +B)y, \quad n\in\N,
\end{equation}
where the Weierstrass function has two periods $2\omega_1$ and $2\omega_2$ which are independent over $\R$. We denote its differential Galois group over $\C(\wp,\dot\wp)$ by $G$. 

Function $v=\wp(t)$ satisfies the differential equation 
\begin{equation}
 \label{eq:el}
{\dot v}^2=4v^3-g_2v-g_3=:f(v) 
\end{equation} 
The algebraic form of Lam\'e equation is obtain from~\eqref{eq:LH} by setting $z=\wp(t)$ and it reads
\begin{equation}
 \label{eq:aLH}
y'' +\frac{1}{2}\frac{f'(z)}{f(z)}y' -\frac{n(n+1)z+B}{f(z)}y=0, \quad n\in\N.
\end{equation} 
Let $G_{\mathrm{AL}} $ be the differential Galois group over $\C(z)$ of this equation. 

As it was show in \cite[Section 5]{Churchill:99::}, $G= G_{\mathrm{AL}}\cap\mathrm{SL}(2,C)$, and moreover it was  also shown that $G$ is finite iff $G_{\mathrm{AL}}$ is finite.  In \cite[Corollary 3.4]{Beukers:04::} it was proved that if $G_{\mathrm{AL}}$ is finite, then $G_{\mathrm{AL}}$ is a dihedral group $D_m$ of order $2m$, for a certain $m\geq3$. In this case, 
\[
 G=D_m\cap  \mathrm{SL}(2,C)=\left\{ \begin{bmatrix}
 \exp 2\pi \mathrm{i}\frac{l}{m} &0\\
0 &\exp -2\pi  \mathrm{i}\frac{l}{m}
\end{bmatrix}
\ |  \ l=0,\ldots,m-1\right\}.
\]
This fact implies that if $G$ is finite, then it is a cyclic group of order $m$ for a certain $m\geq 3$, so  there are two independent solutions $y_1$ and $y_2$  of~\eqref{eq:LH} such that $y^m_i\in\C(\wp,\dot\wp)$, for $i=1,2$.

Now, it is known that for given $n\in\N$,  and $m\geq 3$ the number of linearly non-equivalent Lam\'e equations~\eqref{eq:aLH} with differential Galois $D_m$ is finite, see \cite{Beukers:04::,Dahmen:07::}. 
Nevertheless, for long time it was unclear if there exists a Lam\'e equation~\eqref{eq:aLH} for which $G_{\mathrm{AL}}$ is finite.  This problem, among other things, was analysed by Baldassarri and Dwork, see~\cite{Baldassarri:79::,Baldassarri:80::,Baldassarri:81::}, but only a bound on $m$ was found. Later, see~\cite{Baldassarri:86::,Baldassarri:87::}, examples of Lam\'e equations~\eqref{eq:aLH} with a finite differential Galois group were found.  

In practice, it is important to distinguishing parameters $n$, $B$, $g_2$ and $g_3$ for which $G_{\mathrm{AL}}$ is $D_m$  with prescribed $m\geq 3$.  However, as far as we know, such conditions is difficult to obtain. For $n=1$ and $m=5$ such conditions are given explicitly in~\cite{Baldassarri:86::} where it is conjectured that for arbitrary $n$ and $m$  they should have a polynomial form with respect to variables $B$, $g_2$ and $g_3$.

\end{document}